\documentstyle[12pt]{article}
\begin{document}
\setlength{\baselineskip}{0.30in}

\newcommand{\beq}{\begin{equation}}
\newcommand{\eeq}{\end{equation}}

\newcommand{\bi}{\bibitem}

{\hbox to\hsize{Dec. 1993 \hfill UM-AC-93-31 }}

\begin{center}
\vglue .06in
{\Large \bf {Baryogenesis}}\\[.5in]
{Invited talk at TAUP93\\  LNGS, Italy, September 19-23, 1993} \\
[.1in]
{\bf A.D. Dolgov \footnote{Permanent address: ITEP, 113259, Moscow,
Russia.} }
\\[.05in]
{\it{The Randall Laboratory of Physics\\
University of Michigan, Ann Arbor, MI 48109-1120}}\\[.15in]

{Abstract}\\[-.1in]
\end{center}
\begin{quotation}
A review on the present state of the baryogenesis is given with an emphasis
on electroweak baryogenesis. Technical details of the numerous models
considered in the literature are not elaborated but unresolved  problems
of the isssue are considered. Different logically possible alternatives of
the electroweak scenarios are presented. A possible impact of baryogenesis
on the universe structure formation is discussed.

\end{quotation}

\newpage

Baryogenesis is a process of generation of an excess of baryons over
antibaryons which presumably took place at an early stage of the Universe
evolution. Two questions immediately arise in this connections: first, why
do we need that and, second, if baryogenesis is obligatory or one can make
the observed Universe without it and the existence of baryogenesis at an
early stage (or stages) is only one of several possible alternatives in
cosmology. In my opinion baryogenesis is not only possible and natural
in the frameworks of modern physics but is also necessary for the creation
of the observed Universe at least at the same level as inflation.

The idea of baryogenesis emerged from the observations that the Universe
at some distance scale $l_B$ around us is practically
100\% charge asymmetric with baryon number
density very much exceeding that of antibaryons, $N_B \gg N_{\bar B}$. The
magnitude of the asymmetry is characterized by the ratio of the baryonic
number density to the number density of photons in cosmic microwave
background radiation:
\beq{
\beta = N_B /N_{\gamma} = 10^{-9} - 10^{-10}
}\label{beta}
\eeq
This small number means in particular that the size of the charge asymmetry
(which is practically 100\% now) was tiny at high temperatures,
$T> \Lambda_{QCD} \approx 100$ MeV. At these temperatures antibaryons were
practically equally abundant in the primeval plasma and correspondingly
$(N_B - N_{\bar B} /(N_B + N_{\bar B}) \approx \beta \ll 1$.
Still this number, though very small, is
not easy to obtain and the main goal of theoretical models is to get this
number as large as possible.

There are three important problems related to the scale of the asymmetry
$l_B$:
\begin{itemize}
\item{1.} What is the magnitude of $l_B$? Is it infinite or,
what is practically the same, larger that the present-day horizon,
$l_B > l_U \approx 10^{10}$ years? May it be rather small, say, like
$a few\times 10$ Mpc? In the first case the whole Universe or at least
the visible part is baryon dominated while in the second case there may
be a considerable amount of antibaryons which can be in  principle observed
by their interaction with matter on the boundaries. Still since the
distance is fairly large the gamma-flux from the annihilation would be
sufficiently low.
\item{2.} May the Universe be charge asymmetric only in our neighbourhood,
never mind how large it is (even larger than the horizon), and be
charge symmetric as a whole?  The last possibility is aesthetically
appealing since particle-antiparticle symmetry is restored on large.
\item{3.} Is the amplitude of the asymmetry $\beta$ a constant or may it
be a function of space points $\beta =\beta (x,y,z)$? The last case
corresponds to the so called isocurvature density fluctuations which
may be very interesting for the structure formation in low $\Omega$
Universe.
\end{itemize}

The idea that the dominance of baryons over antibaryons can be explained
dynamically was first proposed by Sakharov  \cite{sakh} in 1967. Before
it was a common belief that a nonzero baryonic charge of the Universe is a
result of mysterious initial conditions. At the present day there are
several hundred papers on the subject discussing different possible
scenarios of the generation of the baryon asymmetry of the Universe. The
history of the problem as well as long lists of references can be found
in the review papers \cite{dz,kt,dad,ckn}. Three very well known by now
conditions of the baryogenesis which were formulated by Sakharov are the
following:
\begin{itemize}
\item{1.} Baryonic charge nonconservation.
\item{2.} Asymmetry in particle-antiparticle interactions (breaking of
C- and CP-invariance).
\item{3.} Deviation from thermal equilibrium.
\end{itemize}

It can be shown that neither of these conditions are obligatory (see
ref. \cite{dad}) but counterexample are rather exotic. A nice feature
of these three conditions is that they are perfectly natural in the
frameworks of the present-day particle physics. Baryonic charge
nonconservation, which was the most problematic 25 years ago, now is
predicted by grand unification models and what's more by the standard
electroweak theory. Unfortunately these are only theoretical arguments
and the proton remains stable despite very strong efforts
to discover its decays. The only "experimental" evidence in favor of
baryonic charge nonconservation is given now by cosmology. On the contrary
C- and CP-violation are observed experimentally in particle physics and
we may be sure that particles and antiparticles are indeed different.
Still theoretically this phenomenon is not well understood: there are
many models for CP-violation and we do not yet know which one is true.
As for deviation from thermal equilibrium it is provided by the universe
expansion and always exists for massive particles with the relative
magnitude of the order $(m^2/T^2) (H/\Gamma)$ where $T$ is the temperature
of the primeval plasma, $H$ is the Hubble parameter characterizing the
expansion rate, and $\Gamma$ is the reaction rate. This expression is valid
for $m \leq T$ and is typically rather small.
For $m>T$ the contribution of the particles with the mass
$m$ is usually exponentially suppressed so in both cases deviations from
equilibrium are small. This smallness is not crucial for scenarios
of baryogenesis at grand unification scale but may be very important for
lower temperatures. Fortunately there is another way to break the
equilibrium by the first order phase transition.
In that case one may expect a low energy baryogenesis,
$T\ll T_{GUT} \approx 10^{16}$ GeV. Anyhow
some deviations from thermal equilibrium always exist in the cosmological
plasma and this provides the third necessary condition for baryogenesis.

We see that baryogenesis might happen in the course of the Universe
evolution and now I would like to argue that it indeed took place. The
crucial point is that inflation is impossible without baryogenesis.
One may argue that the existence of inflation could also be questioned.
Strictly speaking this is true since we do not have rigorous proof that
the Universe, as we see it, cannot be created without inflation. Moreover
this proof can never be presented. However inflation is the only scenario
which solves in a simple way many cosmological problems which cannot be
addressed in any other known cosmological scenario. Among them are the
problems of
\begin{itemize}
\item{1)} flatness; the Universe should be flat with the accuracy $10^{-15}$
during primordial nucleosynthesis,
\item{2)} horizon, homogeneity, and isotropy,
\item{3)} generation of the primordial density fluctuations,
\item{4)} initial push which gave rise to the Universe expansion; the
inflationary equation of state $p=-\rho$ corresponds to antigravitating
medium creating expansion.
\end{itemize}

Of course inflationary models have their own problems like very small
strength of the inflaton interactions and the absence of a natural
inflationary scenario in the frameworks of the
simplest gauge theories of particle interactions which is an argument
against inflation.
On the other hand the prediction of inflationary models
of approximately flat
spectrum of density perturbations is in a reasonable agreement with the
COBE data (see the talk by J.Silk at this Conference). Slightly tilted
spectrum of density fluctuations may better describe the Universe structure
formation and fortunately there exist inflationary models which can give
this prediction (see e.g. \cite{abffo}).
Another quantitative
prediction of inflation that the density parameter $\Omega$ is most probably
equal to one may be in agreement with observations but the latter are very
inaccurate and one cannot make a decisive conclusion here. Plenty of people
would be happy if $\Omega$ is considerably smaller than one. In that case
we may not need nonbaryonic dark matter and in view of the recent
announcements by the experimental groups EROS and MACHOS of
possible microlensing events
(see the talks by A. Milsztajn and B. Sadoulet at this Conference) one
may think that all the dark matter in the Universe is baryonic. The claim
that there is some baryonic dark matter in the Universe is supported by
the primordial nucleosynthesis theory which gives
$\Omega_{B} \approx 0.05(H/50km/sec/Mpc)^{-2}$
\cite{dns} while the contribution of the visible baryonic matter
is $\Omega_{B} \approx 0.01$. However purely baryonic universe encounters
serious difficulties in large scale structure formation especially because
of very small fluctuations of the microwave background temperature. From
this point of view nonbaryonic dark matter and large (close to 1) $\Omega$
are very desirable. Taken together with the nice inflationary solution of
the basic cosmological problems this gives a very strong argument in favor
of inflationary scenario.

For successful solution of the flatness and horizon problems duration of
inflationary stage should be sufficiently large, $H_I t_I \geq 65-70$. If
baryonic charge were conserved it would be diluted during inflation by a
huge factor $e^{210}-e^{195}$. Unnatural by itself it does not exclude
initial conditions with a very big baryonic charge density. But nonzero
baryonic charge density implies simultaneously nonzero energy density
associated with it. Inflation could be achieved only if energy density in the
Universe is a constant or slowly varying function of the scale factor $a$.
This is not true for the energy density associated with baryonic
charge, $\rho_B$. It varies
as $1/a^3$ for nonrelativistic particles and as $1/a^4$ for relativistic ones.
{}From the value of $\beta$ (\ref{beta}) we may conclude that
at high temperature stage $\rho_B \approx 10^{-10} \rho_{tot}$. It means
that the total energy density could be approximately constant for
the period not larger than 6 Hubble times which is too little for a
successful inflation. Thus inflation demands nonconservation of baryons.

Historically first papers on baryogenesis which were based on a well defined
particle physics model were done in the frameworks of the
grand unification theories (for the review
and the literature see \cite{dz,kt}). Grand unification models present a
beautiful extension of the minimal standard $SU(3)\times SU(2) \times
U(1)$-model (MSM). A strong indication of the validity of the grand unification
is the crossing of all three gauge coupling constants of supersymmetric
extension of MSM at the same point near $E_{GUT}=10^{16}$ GeV.
It is rather difficult to believe that there are no new particles
and interactions
in the region between electroweak or low energy supersymmetry scale and grand
unification scale but if the essential quantity is the logarithm
of energy the distance between these two scales is not too big and one may hope
that MSM or supersymmetric version of it is the ultimate truth in low energy
physics (up to $E_{GUT}$). One more argument in favor of low energy
supersymmetry is provided by cosmology, namely, if one demands in accordance
with the theory of large scale structure formation that the bulk of matter
in the universe is in the form of cold dark matter and assumes that the
cross-section of the annihilation of the latter is given by $\sigma =
\alpha^2 /m^2$ then the mass $m$ should be in the  region
100 GeV - 1 TeV. It is just the scale of low energy supersymmetry (for
more details see e.g. ref. \cite{kane})

A strong objection against GUT baryogenesis is a low heating temperature
after inflation. It is typically 4-5 orders of magnitude below
$E_{GUT}$. It means that the GUT era possibly did not exist in the early
Universe. A very interesting alternative to the GUT baryogenesis is the
electroweak one (for the review see refs. \cite{dad,ckn}). Electroweak
theory provides all the necessary ingredients for baryogenesis including
baryon nonconservation (see below) so one may hope to get some baryon
asymmetry of the Universe even in the frameworks of the MSM. A very
interesting question is if it is possible to get the right magnitude of
the asymmetry in MSM or  baryogenesis demands an extension of the minimal
model.

One may say in support of the second possibility that
cosmology already demands physics beyond the standard model. It should be
invoked for realization of inflation, for the generation of the primordial
density perturbations, for nonbaryonic dark matter, etc. A drastic
change in the standard physics may be necessary for the solution of the
cosmological term problem. (There is a hope however that it may be solved
by infrared instability of quantum gravity in De Sitter background, see
e.g. refs. \cite{lf,tw}.)
So we have already a strong evidence that there
is physics beyond the standard model and thus baryogenesis should not be
confined to the MSM. Still the possibility of realistic baryogenesis in the
minimal model is extremely appealing and moreover it gives the unique
possibility to express the magnitude of the baryon asymmetry $\beta$
through parameters of the standard  model measured in direct experiments.

Baryonic charge nonconservation in the electroweak theory was discovered
by 't Hooft \cite{thooft}. It is a very striking phenomenon. Classically
baryonic current, as inferred from  the electroweak Lagrangian, is
conserved
\beq{
\partial_\mu j_{baryonic} ^\mu = 0,
} \label{jclass}
\eeq
but the conservation is destroyed by the quantum corrections. The latter
are given by the very  well known chiral anomaly associated with triangle
fermionic loop in external gauge field. The calculation which can be
found in many textbooks gives
\beq{
\partial_\mu j^\mu _{BL} = N_f \left( {g_2^2 \over 32\pi^2} W\tilde W
-{g_1^2\over 32\pi^2} Y \tilde Y \right)
}\label{jquant}
\eeq
Here $N_f$ is the number of fermionic flavors, $g_{1,2}$ are the gauge
coupling constants of $U(1)$ and $SU(2)$ groups, $W$ and $Y$ are
the gauge field strength tensors for $SU(2)$ and $U(1)$ respectively, and
tilde means dual tensor, $\tilde W^{\mu\nu} = \epsilon ^{\mu\nu\alpha\beta}
W_{\alpha\beta} /2$. The products of the gauge field strength $W\tilde W$
and $Y \tilde Y$ can be written as divergences of vector quantities,
\beq{
W\tilde W = \partial _\mu K^\mu_2
}\label{dk2}
\eeq
\beq{
Y \tilde Y =\partial_\mu K^\mu_1
}\label{dk1}
\eeq
where
\beq{
K_1^\mu =\epsilon ^{\mu\nu\alpha\beta} Y_{\nu\alpha} Y_\beta
}\label{k1}
\eeq
\beq{
K_2^\mu =\epsilon ^{\mu\nu\alpha\beta} (W_{\nu\alpha} W_\beta -{1\over 3}
g_2 W_\nu W_\alpha W_\beta )
}\label{k2}
\eeq
Here $Y_\nu$ and $W_\nu$ are gauge field potentials of abelian $U(1)$ and
nonabelian $SU(2)$ groups respectively. Usually total
derivatives are unobservable since they may be integrated by parts and
disappear. This is true for the contribution into $K^\mu$ from the
gauge field strength tensors $Y_{\mu\nu}$ and $W_{\mu\nu}$
which should sufficiently fast vanish
at infinity. However it is not obligatory for the potentials for which
the integral over infinitely separated hypersurface may be nonzero.
Hence for nonabelian groups the current
nonconservation induced by quantum effects becomes observable.

Because of conditions (\ref{jquant}, \ref{dk2}, \ref{k2})
the variation of the baryonic
charge can be written as
\beq{
\Delta B = N_f \Delta N_{CS}
}\label{db}
\eeq
where $N_{CS}$ is the so-called Chern-Simons number characterizing topology
in the gauge field space. It can be written as a space integral of the time
component of the vector $K^{\mu}$:
\beq{
N_{CS} = {g_2^2 \over 32 \pi^2} \int d^3 x K^t_2
}\label{ncs}
\eeq
Though $N_{CS}$ is not a gauge invariant quantity its variation
$\Delta N_{CS} = N_{CS} (t) -N_{CS} (0) $ is.

In vacuum the field strength tensor $W_{\mu\nu}$ should vanish while
the potentials
are not necessarily zero but can be the so called purely gauge potentials:
\beq{
W_{\mu}=-{i\over g_2} U(x)\partial_\mu U^{-1}(x)
}\label{gauge}
\eeq
There may be two classes of gauge transformations keeping $W_{\mu\nu}=0$:
one that does not change $N_{CS}$ and the second that changes $N_{CS}$.
The first one can be realized by a continues transformation of the potentials
while the second cannot. If one tries to change $N_{CS}$ by a continuous
variation of the potentials one has to pass the region where $W_{\mu\nu}$ is
nonzero. It means that vacuum states with different topological charges
$N_{CS}$ are separated by the potential barriers. The probability of the
barrier penetration can be calculated in quasiclassical approximation. The
trajectory in the field space in imaginary time which connects two vacuum
states differing by a unit topological charge is called the instanton. As in
the usual quantum mechanics action evaluated on this trajectory gives the
probability of the barrier penetration \cite{bpst}:
\beq{
\Gamma \sim \exp \left( {4\pi \over \alpha_W } \right) \approx 10^{-170}
}\label{inst}
\eeq
where $\alpha_W = g_2^2 /4\pi$. This number is so small that it is not
necessary to present a preexponential factor.

Expression (\ref{inst}) gives the probability of the baryonic charge
violation at zero energy. We know from quantum mechanics that the
probability of the barrier penetration rises with rising energy. Moreover
in the system with nonzero temperature a particle may classically go over
the barrier with the probability determined by the Boltzmann exponent,
$\exp (-E/T)$. This analogy let one think that a similar phenomenon may
exist in quantum field theory so that the processes with baryonic charge
violation are not suppressed at high temperature. One should not of course
rely very much on this analogy since there may be a
serious difference between
quantum mechanics which is a system with a finite number of degrees of freedom
and quantum field theory which has an infinite (continuous) number of degrees
of freedom. Still in a detailed investigation of this phenomenon convincing
arguments have been found that baryonic charge nonconservation at high
temperature may be strong and that baryogenesis by electroweak processes may
be possible. A good introduction to the theory of the electroweak
$(B+L)$-violation at high temperature can be found in lectures \cite{mcl}.

The first paper where this idea was seriously considered belongs to Kuzmin,
Rubakov, and Shaposhnikov \cite{krs}
(for the earlier papers see ref. \cite{dad}).
They argued that the probability of
baryonic charge nonconservation at nonzero $T$ is determined by the expression
\beq{
\Gamma \sim \exp \left( -{U_{max} \over T} \right)
}\label{sphal}
\eeq
where $U_{max}$ is the potential energy at the saddle point separating
vacua with different topological charges. The field configuration
corresponding to this saddle point is called sphaleron. It was originally
found in ref. \cite{dsn} and later rediscovered in paper \cite{mant}. In
the last paper the relation of this solution to the topology changing
transitions and baryonic charge nonconservation was clearly understood.
Quantum mechanical analogue of the sphaleron is a single point in the
phase space, i.e. the position of particle sitting at the top
of the barrier. The energy of the sphaleron is
\beq{
U_{max} \equiv U(\phi_{sphaleron} (x) ) ={2M_W\over \alpha_W} f\left(
{\lambda \over g^2}\right)
}\label{usp}
\eeq
where $\lambda$ is the self-interaction coupling constant of the Higgs
field, $f$ is a function which can be calculated numerically,
$f= O(1)$, and $M_W$ is the mass of the W-boson. At zero temperature
$2M_W/\alpha_W \approx 10$ TeV. However at high temperatures close to the
electroweak phase transition the Higgs condensate is gradually destroyed and
the height of the barrier decreases together with the
mass of W-boson $M_W^2(T) = M_{0W}^2 (1-T^2/T^2_c)$ \cite{dak,adl}
where $T_c=O(1TeV-100GeV)$ is the critical temperature of the transition.
Thus one may expect that the processes with baryonic charge nonconservation
are indeed unsuppressed at high temperatures.

The situation is not so simple however and there are a few problems which
should be resolved before a definite conclusion can be made. They mostly
stem from the difference between finite dimensional system like quantum
mechanics and infinitely dimensional field theory. The first question is
what is the probability of the processes with the change of topology in
the gauge field space. Such processes proceed in presumably multiparticle
collisions through formation of the classical field configuration with
the coherent scale which is much larger than inverse temperature.
If these processes are not fast enough the sphalerons
may be not in thermal equilibrium
and possibly far below the equilibrium so that
the expression (\ref{sphal}) would not be applicable. At the
present day we do not know a reliable analytical way to
address this problem. Numerical simulation of the analogous problem
made in 1+1 dimensions \cite{gr} showed that the creation
of soliton-antisoliton pairs are indeed fast enough to maintain the
equilibrium value and this is one the strongest arguments in favor of
efficient baryon nonconservation in electroweak processes. However such
processes in one dimensional space may proceed much easier than those in
three space dimensions simply because in $D=1$ the change of topology means
just a jump from one constant value of the Higgs field to another while
in $D=3$ much more fine tuning in every space point is necessary.
Unfortunately numerical simulation in $3+1$ case is much more difficult
and correspondingly much less reliable. So strictly speaking the
probability of the sphaleron transitions is not known and a better
understanding of it is very much desirable though it seems plausible
that they are not too much suppressed so that thermal equilibrium with
respect to the topology changing transitions was achieved in the early
universe.

Another question related to the probability of the processes
with $\Delta B \not= 0$ is what is the entropy of the sphalerons or in
other words what is the preexponential factor in expression (\ref{sphal}).
This factor characterizes the width of the potential near the saddle
point in the directions orthogonal to the trajectory over potential
barrier and was calculated in ref. \cite{amcl}. With this factor taken
into account the probability of electroweak processes with baryonic charge
nonconservation in the phase with broken electroweak symmetry can be
evaluated as
\beq{
{\Gamma_{\Delta B} \over H} = 10^{24} \left( {M_W(T) \over T} \right)^2
e^{-120M_W(T)/T}
}\label{gamma1}
\eeq
where $H$ is the Hubble parameter characterizing the rate of the Universe
expansion.

At temperatures above electroweak phase transition the rate of baryonic
charge nonconservation is given by \cite{amcl,ks}
\beq{
\Gamma_{\Delta B} \approx \alpha_W^4 T
}\label{gamma2}
\eeq
Recall that expressions (\ref{gamma1}) and (\ref{gamma2}) are valid
only if sphalerons are in thermal equilibrium. If this is true then
$\Gamma_{\Delta B} /H \gg 1$ at high temperatures and then abruptly falls
down with falling temperatures. Thus processes with baryonic charge
nonconservation are in equilibrium at high $T$ and at some point are
instantly switched off. Thus any preexisting baryon asymmetry would be
washed out and a new one cannot be generated. This conclusion can be
avoided however if deviations from thermal equilibrium existed at
the time when baryonic charge nonconservation was still effective. This
can be realized in particular if electroweak phase transition is of
the first order. However it is still an open question what is the
type of the phase transition depending in particular on the value of
the Higgs boson mass.

One more comment may be in order here. We spoke before only about baryonic
charge nonconservation. In fact electroweak interactions break equally
baryonic and leptonic charges so that $(B-L)$ is conserved. With this
correction in mind all the previous statements remain true with the
substitution of $(B+L)$ instead of $B$.

Thus the following logical possibilities exist for the
electroweak baryogenesis(we simply enumerate them here and discuss in
some more detail giving recent references below):
\begin{itemize}

\item{I.} Change of the field
topology is suppressed in three-dimensional space.
Sphalerons are never abundant and electroweak nonconservation of $(B+L)$
is ineffective. In that case we should return either to GUT baryogenesis
or to some other more recent proposals described in review paper \cite{dad}.

\item{II.} Sphaleron transitions
are not suppressed above and near the electroweak
phase transition and so $(B+L)$ is strongly nonconserved at these temperatures.
If this is true the following two possibilities are open:

\item{II.1}. The electroweak
phase transition is of the second order and so the
baryon nonconserving processes, which were with a very
good accuracy in thermal
equilibrium above the phase transition, would be
instantly completely switched off
below it. In this case any preexisting $(B+L)$ would be washed out and we
again meet two possibilities:

\item{1a.} The observed
asymmetry might arise from an earlier generated $(B-L)$
either by $(B-L)$ nonconserved processes which exist e.g. in higher rank
grand unification groups or by lepton charge nonconservation in decays of
heavy Majorana fermion.

\item{1b.} Baryogenesis should
take place at low energies below electroweak scale
which for sure demands new low energy weak physics.

\item{II.2.} Electroweak
phase transition is first order so thermal equilibrium was
strongly broken when both phases coexisted. If this is the case
$(B+L)$-asymmetry could be generated in electroweak processes at
temperatures near 1TeV. An important subdivision in this situation is:

\item{2a.} The standard model is able to give
a correct magnitude of the baryon asymmetry
of the Universe so that baryogenesis does not demand any physics beyond the
minimal standard $SU(3)\times SU(2)\times U(1)$-model (MSM).

\item{2b.} An extension of the
minimal standard model is necessary. This is not
well defined and may include an introduction of additional Higgs fields
(like in supersymmetric versions),
considerable CP-violation in the lepton sector, CP-violation in strong
interaction, etc.

\end{itemize}

The essential quantity which determines the character of the phase transition
in the minimal standard model is the magnitude of the Higgs boson mass. For
a large value of the latter the phase transition is second order and for
a small one it is first order. To illustrate this statement let us consider
the following temperature dependent effective potential for the Higgs field
$\phi$ (temperature dependent terms appear due to interactions of the field
$\phi$ with the thermal environment of the cosmic plasma):
\beq{
U(\phi, T) = m^2(T) \phi^2 /2 + (\lambda \phi^4) \ln (\phi^2 /\sigma^2)/4
+\gamma(T_) \phi^3 +...
}\label{UT}
\eeq
Notations here are selfexplanatory. The temperature dependence of the
effective mass is roughly speaking the following $m^2(T) =-m^2_0 +AT^2$
where the constant $A$ is usually positive. (It is positive in MSM.)
Logarithmic dependence on $\phi$ came from one-loop quantum perturbative
corrections to the potential. At
high temperatures the potential has the only minimum at $T=0$, vacuum
expectation value of the $\phi$ is zero, and the electroweak symmetry
is unbroken. At smaller temperatures a deeper minimum is developed at
nonzero $\phi$ and mass of the field near this minimum is $m_H^2
\approx 2m^2_0$ (we neglected here logarithmic terms in $U$).
One sees that the larger is $m^2_0$
(and correspondingly the physical mass $m_H^2$) the easier is the phase
transition. There is no consensus in the literature about the value of $m_H$
separating first and second order phase transitions. While earlier
perturbative calculations in the MSM \cite{mbound} give a rather small value
$m_H\approx 45$ GeV, it was argued that higher loop effects are essential
\cite{dlhll,ae,bd}. Moreover since thermal perturbation theory for nonabelian
gauge fields suffers from severe infrared divergences, nonperturbative effects
might be important acting in favor of the first order phase transition with
higher $m_H$ \cite{ms1}.
It is supported by the recent lattice calculations
\cite{kbis,krs1}. For a more detailed discussion and list of references see
papers \cite{ckn,fs1}. Hence we cannot make any rigorous conclusion now
about the nature of the electroweak phase transition though it seems probable
that MSM with the existing lower experimental bound on the Higgs mass
$m_H>62$ GeV given by LEP favors
second order phase transition while in extended models with several Higgs
fields
the transition might be first order.

Even if the electroweak phase transition in MSM is first order the generated
asymmetry is expected to be very small. It is connected with a strong
suppression of CP-violating effects at high temperatures. CP-breaking in
the MSM is created by the imaginary part of the quark mass matrix
(Cabibbo-Kobayashi-Maskawa matrix).
If there are only two quark generations the imaginary
part is not observable because the phase may be absorbed in a redefinition of
the quark wave function. The statement remains true with more
quarks families with degenerate masses because the unit matrix is
invariant with respect to unitary transformations. One can see that the
minimum number of quark families for which the imaginary part is observable is
three with different masses of quarks with the same value of electric charge.
(If we believe that there is no extension of the standard model then the
necessity of CP-violation for the generation of the charge asymmetry of
the Universe justifies the existence of at least three fermionic families.)
Moreover the amplitude of CP-violation is proportional to the mixing angles
between different families because if the quark mass matrix and the kinetic
term in the Lagrangian are simultaneously diagonal then the phase rotation
would not change them. By these reasons the amplitude of CP-violation in  MSM
is suppressed by the factor (which is called the Jarlskog determinant):
\begin{eqnarray}
A_- &\sim & \sin \theta_{12} \sin \theta_{23} \sin \theta_{31} \sin \delta_{CP}
(m_t^2 -m_u^2)(m_t^2-m_c^2)
\nonumber \\
& &
(m_c^2-m_u^2)(m_b^2-m_s^2)(m_b^2-m_d^2)
(m_s^2-m_d^2) /E^{12}
\label{jdet}
\end{eqnarray}
Here $\theta_{ij}$ are mixing angles between different generations and
$\delta_{CP}$ is the CP-odd phase in the mass matrix. The product of $\sin$'s
of these quantities is about $10^{-4}-10^{-5}$. $E$ is the characteristic
energy of a process with CP-breaking. In the case considered when the
temperature of the medium is above 100 GeV, $E$ is of the same order of
magnitude. Correspondingly one should expect that baryon asymmetry in MSM
should be of the order of $10^{-20}$.

This conclusion was questioned recently by Farrar and Shaposhnikov
\cite{fs1,fs2}.
They argued that flavor dependent temperature corrections to the
quark masses in the vicinity of the domain wall where the expectation
value of the Higgs field is changing nonadiabatically,
may drastically enhance efficiency of the electroweak baryogenesis.
This effect is especially pronounced at
the low energy tail of the quark distribution in the phase space.
As a result
the value of the baryon asymmetry may be close to the observed one even in
the minimal standard model.
This very interesting proposal is discussed by Shaposhnikov at this
Conference so I would not stop on the details of the model.

Despite all the attractiveness of the possibility of effective baryogenesis
in the MSM it should be excluded if the experimental lower bound on the Higgs
boson mass proves to be above the value necessary for successful first
order phase transition. This seems rather probable now and the models with
several Higgs fields are possibly the next best choice. They may give a
larger CP-violation and what's more in these models
both experimental and theoretical bounds
on the Higgs boson mass are much less restrictive.

The generic feature of all scenarios of electroweak baryogenesis is a
coexistence of two phases in one of which baryonic charge is strongly
nonconserved, the corresponding reactions are well in equilibrium, and
no asymmetry can be generated,
while in the second phase baryonic charge is practically conserved
and the asymmetry also cannot be generated though by an opposite reason.
So the only place where baryon asymmetry may be produced are the
boundaries between the phases. The outcome of such a process strongly depends
upon the interaction between the high temperature cosmic plasma and the
domain walls and in particular upon the
velocity of the wall propagation in plasma. These problems are addressed
in several papers (for the recent ones see e.g. refs. \cite{mlmv1,mlmv2}) but
still more work in this field is desirable.

Despite all these uncertainties the electroweak baryogenesis is presently
the most fashionable scenario of creation of the building blocks of our
Universe. There is a large selection of models the literature, each having
a chance to be the right one. A possible exception is the model with a large
CP-violation in the lepton sector \cite{ckn2} which demands a heavy
tau-neutrino with the mass of the order of 10 MeV. However the recent
nucleosynthesis bounds \cite{ktcs,dr,kkkssw} which close
the window for $\nu_\tau$-mass in the region 0.5-35 MeV strongly disfavor it.

In the case if the phase transition is second order, baryon asymmetry could
not be generated by electroweak processes but, if sphalerons are effective,
the latter may be very good for erasure of any preexisting $(B+L)$-asymmetry.
A nonzero initial $(B-L)_i$-asymmetry is conserved by electroweak interactions
and the subsequent sphaleron processes would result in equal baryon and lepton
asymmetry $B_f=L_f=(B-L)_i/2$.
Assuming that this is indeed the case one can derive a bound on the strength
of $(B-L)$-nonconserving interactions at lower temperatures when (and if)
$(B+L)$-erasure is effective. (One should keep in mind however that all these
bounds are valid only if there is no baryogenesis at electroweak or lower
temperature range.) If the rate of $(B+L)$-nonconserving sphaleron
transitions is given by eqs.(\ref{gamma1}, \ref{gamma2}),
the sphaleron processes are
in equilibrium in the temperature range
\beq{
10^2-10^3 <T<10^{12}\, (GeV)
}\label{eqsp}
\eeq
For successful baryogenesis the processes with $(B-L)$-nonconservation
should not be in equilibrium in this range. This idea was first used in
ref.\cite{fy}, where the model of baryogenesis through the decay of heavy
Majorana fermion has been proposed, to put a bound on the Majorana mass
of light neutrinos, $m_M(\nu) <50$ KeV. Neutrinos with a larger Majorana
mass together with sphalerons would destroy both baryon and lepton asymmetry.
There exists a large literature on the subject (the references can be found
in the review paper \cite{dad}) where the bounds on different types of
$(B-L)$-nonconserving interactions are obtained. I would like to mention
here only a recent paper \cite{cko}
where it was argued that lepton asymmetry stored
in right-handed electrons, which are singlets with respect to nonabelian
part of the electroweak group and due to that do not interact with sphalerons,
might be preserved for the temperature down to approximately
10 TeV. Below that the Higgs bosons would effectively transform
right-handed electrons into left-handed ones and subsequently sphalerons
would convert the lepton asymmetry in the sector of right-handed electrons
into baryon asymmetry. The creation of the initial lepton  asymmetry
could be favored by a
rather strong violation of leptonic charge conservation.
All other $(B-L)$-breaking interactions should be
out of equilibrium above 10 TeV while the usual demand is that they are
out of equilibrium at much higher temperatures where either sphalerons
come into equilibrium or where the initial $(B-L)$ is produced.
This invalidates some of the conclusions obtained in the earlier papers (not
quoted here) of stronger bounds on $(B-L)$-nonconservation.
Still the assumption that baryogenesis proceeds through transformation
of an initial $(B-L)$-asymmetry into B-asymmetry permits
to deduce in some cases
more interesting bounds on e.g. L-nonconservation than that following from
direct experiments. There are too many possible forms of the interaction
and theoretical models giving rise to them so that their more detailed
description is outside the scope of the present talk and
one should be addressed to original literature on the subject.

Now I would like to turn to some more exotic cases. The first one is a
possibility of a large lepton asymmetry together with a normal small
baryon asymmetry. Though the data gives a rather accurate value of $\beta$
(within an order of magnitude), the value of the lepton asymmetry is
practically unknown. The best limits follow from the primordial
nucleosynthesis which permits muonic and taonic lepton asymmetry close to
unity while electronic lepton asymmetry cannot exceed 1\%
(see \cite{dad} for the list of references). The bound on
the  chemical potential associated with electronic charge
is stronger because it would directly
shift proton-neutron equilibrium in weak reactions like $n+\nu_e
\leftrightarrow p+e^-$, while $\nu_\mu$ and $\nu_\tau$ influence
$n/p$-ratio only through the total energy density. Thus even in the most
restricted case the value of lepton asymmetry may be as large as
$10^{-2}$.

A large lepton asymmetry could only be realized if the sphaleron processes
were not effective or if the asymmetry was generated below electroweak scale.
Even if this is true,
the  majority of models naturally give $L\approx B$ but there
are some examples permitting $L\gg B$ (see e.g. \cite{dk,dad}). In this case
we would have at our disposal an extra free parameter for the theory of
primordial nucleosynthesis, namely the chemical potential of leptons. What's
more the characteristic scale of spatial variation of the leptonic charge
density $l_L$ might be much smaller than $l_B$ and if the former is in the
range $l_{gal} < l_L <l_U$ one may observe that by spatial variation of the
abundances of light nuclei and in particular of $^4 He$.

The relatively strong isocurvature fluctuations in leptonic sector with a
possibly nonflat spectrum may be also interesting for the theory of the large
scale structure formation
with a single dominant component of hot dark matter.
Usually one considers isocurvature perturbations in baryonic sector which are
stronger bounded by the isotropy of the cosmic microwave background.

Returning to the isocurvature fluctuations in baryonic sector one
may find plenty baryogenesis scenarios (see \cite{dad})
providing very interesting
perturbations with the spectrum varying from the flat one to that having a
prominent peak at a particular wave length. The last case corresponds to
a periodic in space distribution of baryonic matter. It may be naturally
realized if three rather innocent assumptions are satisfied:
\begin{itemize}
\item{1.}There exists a complex scalar field $\phi$ with the mass which is
small in comparison with the Hubble parameter during inflation. The latter
may be as large as $10^{14}$ GeV so one does not need a really light scalar
field.
\item{2.} The potential of the field $\phi$ contains nonharmonic terms like
$\lambda |\phi|^4$.
\item{3.}A condensate of $\phi$ was formed during inflationary stage which
was a slowly varying function of space points.
\end{itemize}
If these conditions are fulfilled then it can be proven (for the details
see refs. \cite{dikn,cd} that the distribution of baryons in the Universe
would be in the form:
\beq{
N_B \approx N_{B0} + N_1 \cos {{\vec r} {\vec n} \over l_B}
}\label{period}
\eeq
where $\vec n$ is an arbitrary unit vector. The scale $l_B$ of the
fluctuations is given by the exponentially stretched Compton wave length of
$\phi$ and could easily be as large as 100 Mpc as was indicated by the
observations \cite{beks}. An interesting picture emerges if $N_0=0$
and the Universe consists of alternating baryonic and antibaryonic layers.

Another unusual picture of the Universe, the so called island universe model
may be realized with the specific though not too complicated model of
baryogenesis \cite{dikn}. In this model our Universe is a huge baryonic
island with the size large or about $10^{10}$ years (or $z=5-10$), while
floating in the see of dark matter which is more or less uniformly distributed.
There are two interesting features of this model which may be relevant to the
structure formation. First, the background radiation comes to us from the
baryon empty regions so that the fluctuations in its temperature is not
directly related to the density perturbations inside the island. Second,
our noncentral position inside the island would give rise to intrinsic
dipole, $d\sim 10^{-3}$, in the angular distribution of the microwave
radiation which is not related to our motion. The quadrupole asymmetry in
this case would be rather small, $q\sim d^2 \sim 10^{-6}$. It may make
easier structure formation in the cold dark matter model. (This point
was emphasized to me by J. Silk.) Without intrinsic dipole and with the
flat spectrum of perturbations more complicated models of the structure
formation are necessary, like e.g. a mixture of hot and cold dark matter
\cite{jp} or a model with cold dark matter and nonzero vacuum energy
(cosmological constant). Both these models demand some fine tuning which
is not well understood today. The first one needs the energy density of
hot and cold dark matter to be the same within the factor of 2 while the
other demands $\rho_{vac}$ which is normally time independent constant
to be close today to the critical energy density which is time dependent,
$\rho_c \sim m_{Pl}^2/t^2$. The latter may be explained if the smallness
of the cosmological constant is ensured by the so called adjustment
mechanism (for the review see \cite{sw}). Though these two possibilities are
more conservative than the island model still they are not the most economic
ones. Proliferation of the universe components from the purely baryonic
universe to the mixed baryonic and hot dark matter or later on to baryonic
and cold dark matter and now to the mixture of all three of them
(baryonic+cold+hot)
with close energy densities is rather mysterious. On the other hand
there are stable neutrinos which are very likely to be massive and it is
also very plausible that there is supersymmetry in particle physics so that
there should be a stable heavy particle. These two are perfect candidates
for the hot and cold dark matter (what's more we may have now dark solar
size objects in galaxies) so that it would be only natural that these particles
participates as building blocks of the Universe. The unresolved question is
their interaction strength which provides very different number densities
and similar mass densities for the particles of hot and cold dark matter.

One may try to make a cosmological model assuming that the only massive
stable particles in the Universe are protons and electrons \cite{ds} and
all the dark matter is made of the normal baryonic staff. To
do that one has to develop a scenario in which baryogenesis proceed much
more efficiently in relatively small space regions giving $\beta =1-0.01$
while it goes normally outside. The regions with that huge baryon number
density mostly form black holes with the mass distribution
\beq{
{dN\over dM} \sim  \exp \left(-\gamma \ln^2 {M\over m_0} \right)
}\label{bhmass}
\eeq
Parameters $\gamma$ and $M_0$ cannot reliably found in the model but one
reasonably expect that $\gamma =O(1)$ and $M_0$ is close to the solar mass.
These black holes might be the objects observed in the microlensing
observations
reported here. If there are no other massive stable particles one has to
build a theory of the structure formation with these black holes which
behave as normal cold dark matter. At the tail of the distribution in mass
there should be very heavy black holes with masses like $10^6-10^9$ solar
masses which may serve as seeds for the structure formation. Still tilted
spectrum of the initial perturbations may be desirable if only cold dark
matter is permitted.

{\bf Conclusions} of the talk reflect to a large extend my personal opinion
and may not be shared by everybody or not even by the majority.
\begin{itemize}
\item{I.} The best choice for the baryogenesis scenario is the electroweak
one and in its framework the one based on the minimal standard model is
the most appealing. The problems with the electroweak baryogenesis are
the unknown probabilities of three dimensional reactions with classical
field configurations, which may question the scenario as a whole, and
the type of the electroweak phase transition. The knowledge of the value
of the Higgs boson mass could be of great help here.
\item{II.} If not MSM the low energy SUSY is the next best choice. SSC could
be very interesting for that but alas...
\item{III.} If electroweak interactions destroy but not generate baryon
asymmetry (like e.g. in the case of the second order phase transition),
a very interesting possibility is baryogenesis through leptogenesis. One
needs to this end a heavy Majorana fermion with mass around $10^{12}$ Gev
(plus-minus a few orders of magnitude) and correspondingly a new physics
beyond the standard model.
\item{IV.} A very low temperature (below the electroweak scale) baryogenesis
is not excluded but there is no natural particle physics model for that.
\item{V.} Majority of models give lepton and baryon asymmetry of
approximately the same magnitude but one may find scenarios giving
$L\gg B$ with interesting consequences for the primordial nucleosynthesis.
\item{VI.} A better understanding of baryogenesis may be of interest for
the theory of the large scale structure formation in particular because
in the process of baryogenesis isocurvature density fluctuations
with a complicated spectrum might be created.
\end{itemize}

This work was supported in part by the Department of Physics of University
of Michigan and by NSF Young Investigator reward to F. Adams.


\end{document}